\documentclass[12pt]{iopart}
\usepackage{graphicx,wrapfig}

\newcommand{\vtwo}{v_2}
\newcommand{\vtwoC}{v_{2c}}
\newcommand{\Raa}{R_{AA}}
\newcommand{\pT}{p_T}
\newcommand{\cosTwoPhi}{ cos\left(2[\phi-\psi_{RP}]\right) }
\newcommand{\mean}[1] { \left< #1 \right>}

\begin{document}
% Journal identifier can be put here if required, e.g.
%\jl{14}

\title{Single Electron Elliptic Flow Measurements in Au+Au collisions from STAR}

\author{Frank Laue\dag  (for the STAR Collaboration)}

\ead{laue@bnl.gov}
\address{\dag\ Brookhaven National Laboratory,  Bldg 510A, Upton, NY-11973, USA}

\begin{abstract}
Recent measurements of elliptic flow $(v_{2})$ and the nuclear modification
factor $(R_{CP})$ of strange mesons and baryons in the intermediate $p_{T}$
domain in Au+Au collisions demonstrate a scaling with the
number-of-constituent-quarks.  This suggests hadron production via quark
coalescence from a thermalized parton system. Measuring the elliptic flow of
charmed hadrons, which are  believed to originate rather from
fragmentation than from coalescence processes, might therefore change our
view of hadron production in heavy ion collisions.

While direct $v_{2}$ measurements of charmed hadrons are currently not
available, single electron $v_{2}$ at sufficiently high transverse momenta
can serve as a substitute. At transverse momenta above $2$~GeV/c, the
production of single electrons from non-photonic sources is expected to be
dominated by the decay of charmed hadrons. Simulations show a strong
correlation between the flow of the charmed hadrons and the flow of their
decay electrons for $p_T > 2$~GeV/c.

We will present preliminary STAR results from our single electron $v_2$ measurements from
Au+Au collisions at RHIC energies.

\end{abstract}


%\submitted

\maketitle

\section{Introduction}
\begin{wrapfigure}{r}{10cm}
\begin{center}
\includegraphics[width=10cm]{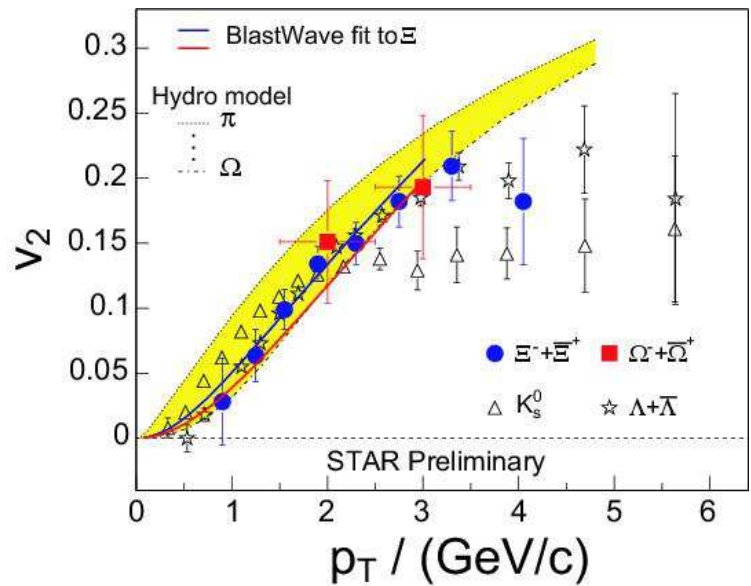}
\end{center}
\caption{Elliptic flow amplitude $\vtwo$ for various strange hadrons as a function of transverse momentum $\pT$. 
%The yellow band represents the $\vtwo$--expectations from hydro-dynamic models, where the upper and lower boundaries reflect the range given by the pion and Omega mass
}
\label{fig:strangeV2}
%\end{figure}
\end{wrapfigure}

Elliptic flow of hadrons has long been suggested as a sensitive probe of the pressure built up in the early stages of heavy ion collisions \cite{Ollitraut:92}. It has furthermore been proposed as a signal of a phase transition from Quark-Gluon-Plasma to hadronic matter \cite{Sorge:99}. At RHIC energies, for the first time in relativistic heavy ion collisions, elliptic flow of unflavored and strange hadrons\footnote{Throughout this manuscript we will call the $u$ and $d$ quarks the {\it unflavored} quarks. Hadrons built up exclusively from {\it unflavored}  constituent quarks will be called {\it unflavored} hadrons. In this notation, the $\phi$-meson will be {\it flavored}. $u,d$ and $s$ quarks will be called {\it light} quarks.} was found to be reasonably well described by hydro-dynamic calculations \cite{HuovinenPhysLettB503_58} up to transverse momenta of at least $\pT=1.5$~GeV/c as shown in Figure \ref{fig:strangeV2}\cite{StarJPhysG30_S693} for the 80\% most central events from $Au+Au$ collisions at $\sqrt{s_{NN}}=200$~GeV.

At transverse momenta above $\pT=2$~GeV/c however, the measured $\vtwo$ values seem to saturate and therefore are not describe by the hydro-dynamic calculations anymore. More interesting than this deviation between data and calculation is an apparent splitting between mesons and baryons, independent of their masses. While mass ordering is an essential feature of hydro-dynamics and has been shown to lead to lower $\vtwo$ values for heavier particles in the low transverse momentum region \cite{StarPhysRevLett87_182301}, a mass independent splitting between mesons and baryons has also been observed in the nuclear modification factor $R_{AA}$ \cite{StarPhysRevLett92_052303}.  Voloshin first suggested a scaling of $\vtwo$ and $\Raa$ with the number of constituent quarks for all particle species \cite{VoloshinNuclPhysA715},  which has been interpreted as a signal for a partonic phase in the early stages of the collision. This interpretation is supported by quark coalescence and quark recombination models, which are able to reproduce the measured $\vtwo$ and $\Raa$ values at intermediate transverse momenta ($\pT \approx 2$---$6$~Gev/c) when assuming a system of thermalized quarks \cite{Nonaka:2004}.  Please note, at even higher transverse momenta ($\pT>6$~GeV/c) , hadron production is expected to be dominated by parton fragmentation, hence quark coalesence models should break down in the high-$\pT$ region \cite{Molnar:2004}. 

In the regime where quark coalescence is valid, the elliptic flow for all particle species can be expressed as
\begin{equation}
v_{2h} = \sum_{i=0}^n v_{2q_i} \left(\frac{1}{n} \pT \right) \ \ \  \cite{Nonaka:2004,Molnar:2004}. 
\label{equ:scale}
\end{equation}
 Here,  $n$ is the number of constituent quarks  and $v_{2h}$ ($v_{2q_i}$)  is the hadron (quark) elliptic flow amplitude.
 
\begin{figure}[htp]
\begin{center}
\includegraphics[width=10cm]{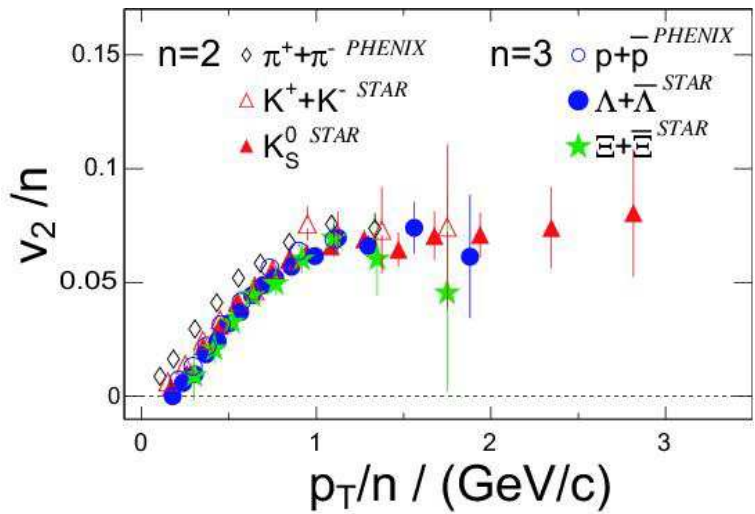}
\end{center}
\caption{Scaled elliptic flow amplitudes $\vtwo/n$ as a function of $\pT/n$ for mesons and baryons.}
\label{fig:scaledV2}
\end{figure}

Figure \ref{fig:scaledV2} \cite{StarJPhysG30_S693} shows the scaled flow amplitudes  $v_2/n$ as a function of $\pT /n$ for three meson and three baryon species,  again for the 80\% most central events from $Au+Au$ collisions at $\sqrt{s_{NN}}=200$~GeV.  Ignoring the low momentum pions, which suffer from strong resonance feed-down,  the number-of-constituent-quark-scaling seems to nicely hold for hadrons built up from $u,d$ and $s$ quarks.  Note, there is no evident ordering between hadrons of different strangeness content, suggesting that the $s$ quark's flow is similar to that of the $u,d$ quarks \cite{Nonaka:2004} .  
   
With the arrival of the first charm measurements from the RHIC detectors, it is now of great interest  to study the elliptic flow of charmed hadrons.  Within quark coalescence models, and $\vtwo$ of $u,d$ and $s$ quarks being determined from Figures \ref{fig:strangeV2} and \ref{fig:scaledV2}, elliptic flow measurements from D-mesons present a direct measure of the $c$ quark elliptic flow ($\vtwoC$).  In contrast to the $u,d$ and $s$ quarks which are predominantly produced in soft collisions, $c$ quarks are expected to be produced via hard processes. At the time of their creation, $c$ quarks should therefore be unaware of the collision geometry, i.e. carry no azimuthal anisotropies. Due to the large mass of the $c$ quark, Dong \etal \cite{Dong:2004}  argue, that large $\vtwoC $ values can only be achieved if re-scattering in a partonic phase is strong enough to also thermalized the light quarks. 

The measurement of D-meson elliptic flow might therefore not only challenge our current understanding of charmed hadron production, but also strengthen the case of  a system with partonic degrees of freedom and is a sensitive probe of thermalization therein.
 
\section{Concepts}
\begin{figure}[htp]
\begin{center}
\includegraphics[width=7cm]{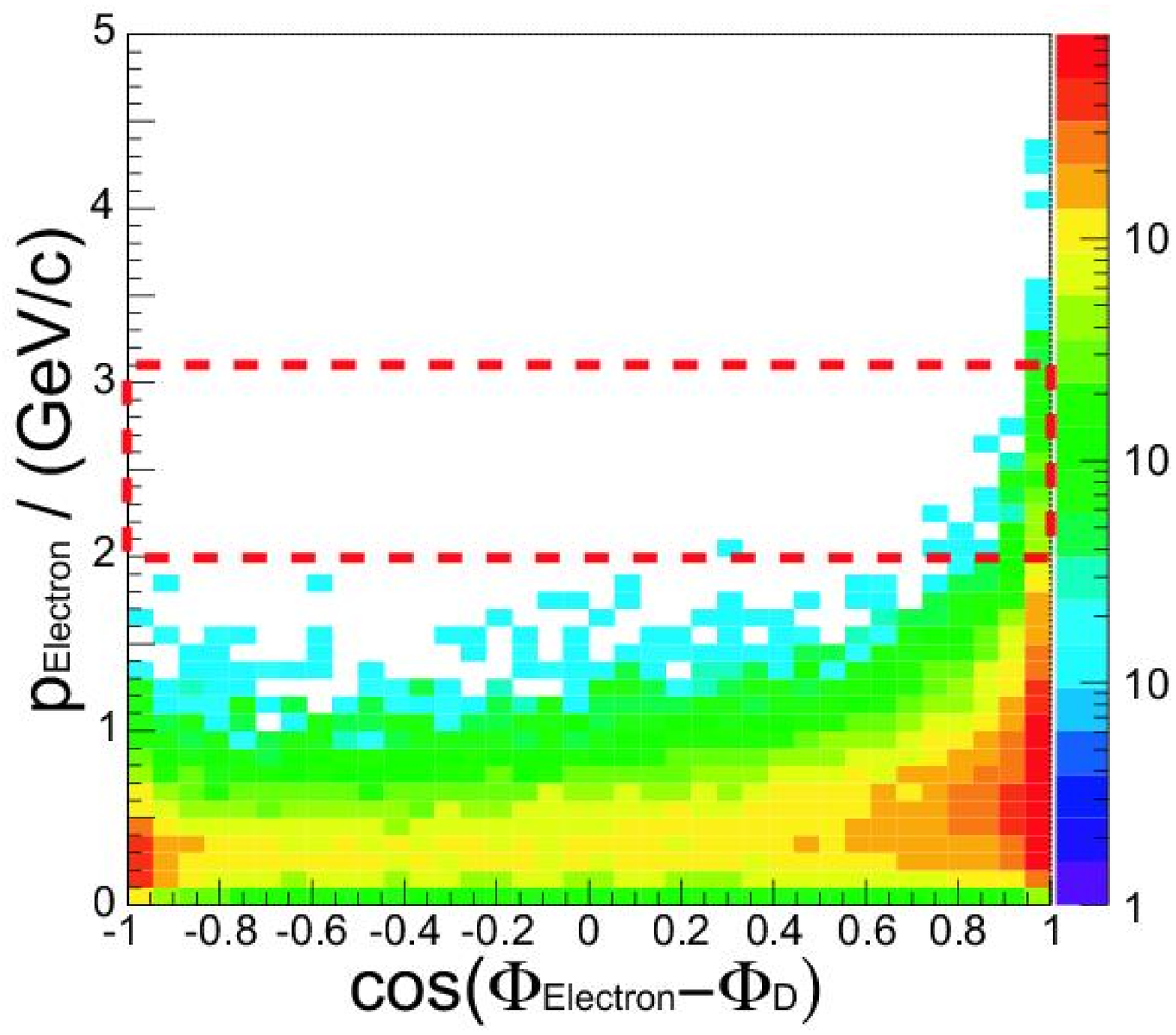}
\includegraphics[width=8cm,height=6cm]{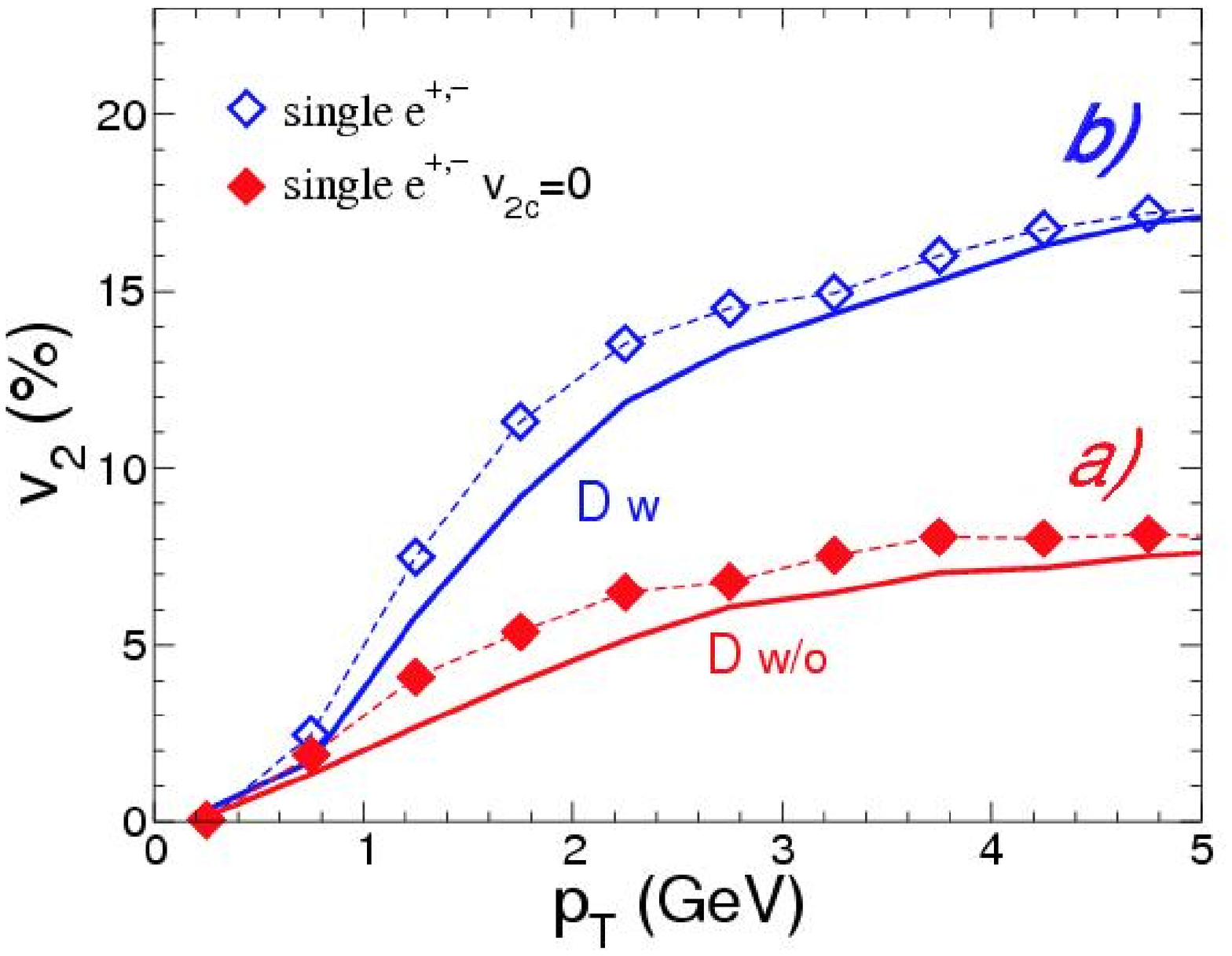}
\end{center}
\caption{(Color online) Left: Correlation between the difference of the D-meson's and its decay electrons's azimuth  emission angles ($\phi_{Electron}-\phi_D$) and  the electron's momentum. Right: Elliptic flow of D-mesons (solid lines) and of single electrons from their decays (dashed lines, symbols). Figure taken from \cite{GrecoNuclTh0405040}.  }
\label{fig:greco_v2_charm}
\end{figure}

Although we have reported on measurements of open charm directly via invariant mass reconstruction  of D-Mesons from hadronic decays, i.e. $D\rightarrow K+\pi$,  in d+Au collisions \cite{StarOpenCharmPaper}, low signal statistic and the large combinatoric background beneath the signal do not allow to study azimuthal anisotropies. However,  at higher transverse momenta semileptonic decays ($D\rightarrow e + X$) are expected to dominate the single electron spectrum and for  momenta higher than $p>2$~GeV/c,  the decay electron's azimuthal angle reflects the D-meson's azimuth angle very accurately as shown in the left panel of Figure \ref{fig:greco_v2_charm}. Plotted is the cosine of the difference of the D-meson's and its decay electron's azimuthal emission angles ($\phi_{Electron}-\phi_{D}$) is versus the electron momentum. The dashed box reflects the momentum region used for this manuscript. Simulations done for this analysis and in \cite{Dong:2004, GrecoNuclTh0312100, KanetaNuclEx0404014}  suggests that single electron $v_2$ is a good measure of open charm flow as seen in the right panel of Figure \ref{fig:greco_v2_charm} (taken from \cite{GrecoNuclTh0405040}). Here, the solid lines depict the D-meson flow from quark coalescence under the assumptions that the $c$ quark a) does not flow ($v_{2c}=0$, lower line), and b) flows just as the light quarks ($v_{2c} = v_{2q-light}$, upper line).   The dashed lines and  the symbols represent the flow of single electrons from D-meson decays under these assumptions. 

Note, for both assumptions, only D-meson production from coalescence is taken into account. Possible anisotropies from jet fragmentation are omitted.

\section{Analysis}
For the analysis reported on here, we studied data from Au+Au collisions at  $\sqrt{s_{NN}}=200$~GeV  recorded by STAR\cite{Star} during Brookhaven's RHIC beam-time in the year 2001. Selecting the 80\% most central events from a minimum bias triggered data set and after applying quality cuts (e.g. requiring  the primary vertex to be located within $\pm35$~cm from the center of the STAR {\bf T}ime {\bf P}rotection {\bf C}hamber (TPC) \cite{StarTPC} along the beam-axis), we used about 2M events. 

\begin{figure}[htp]
\begin{center}
\includegraphics[height=6.5cm]{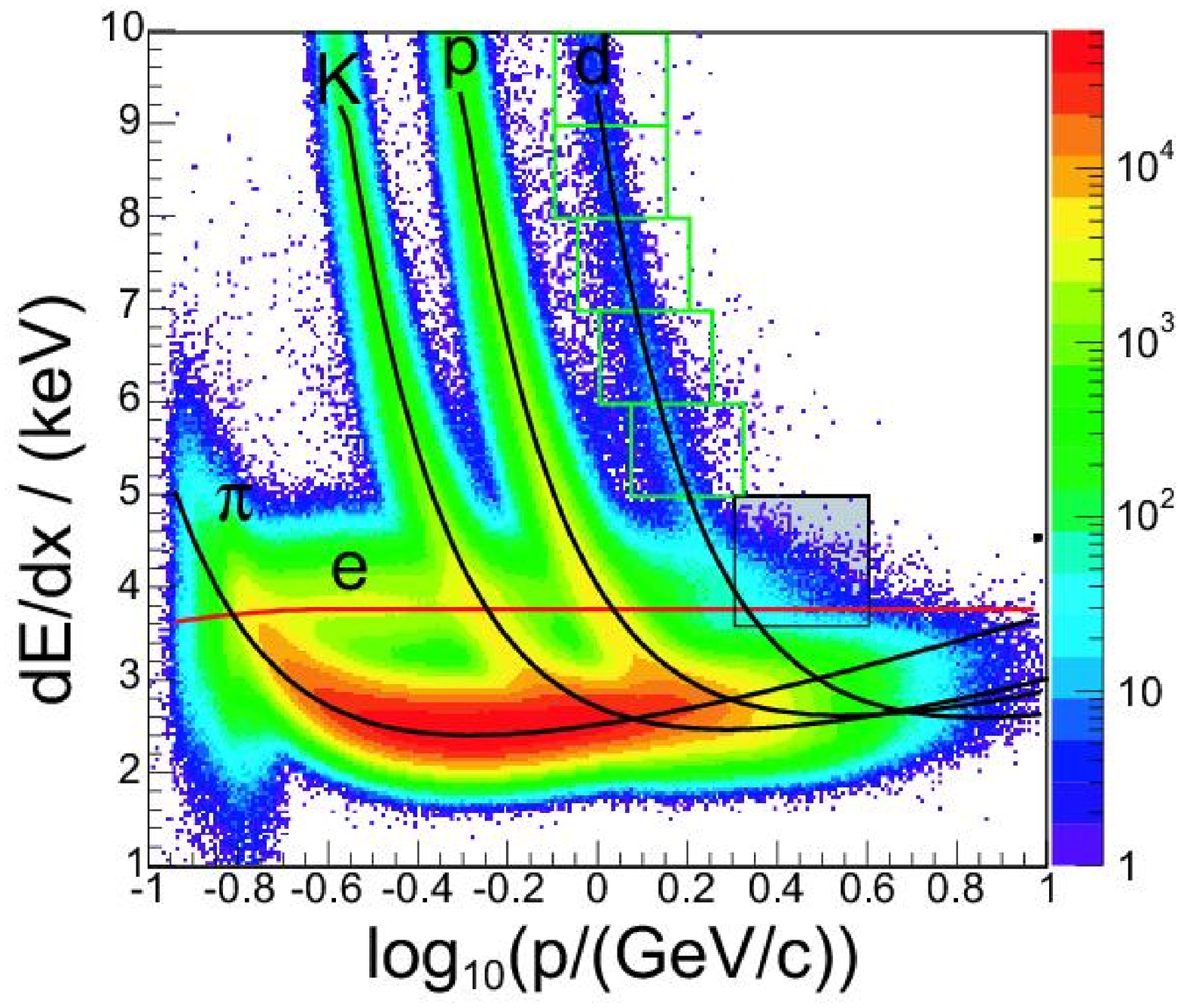}
\includegraphics[height=6cm]{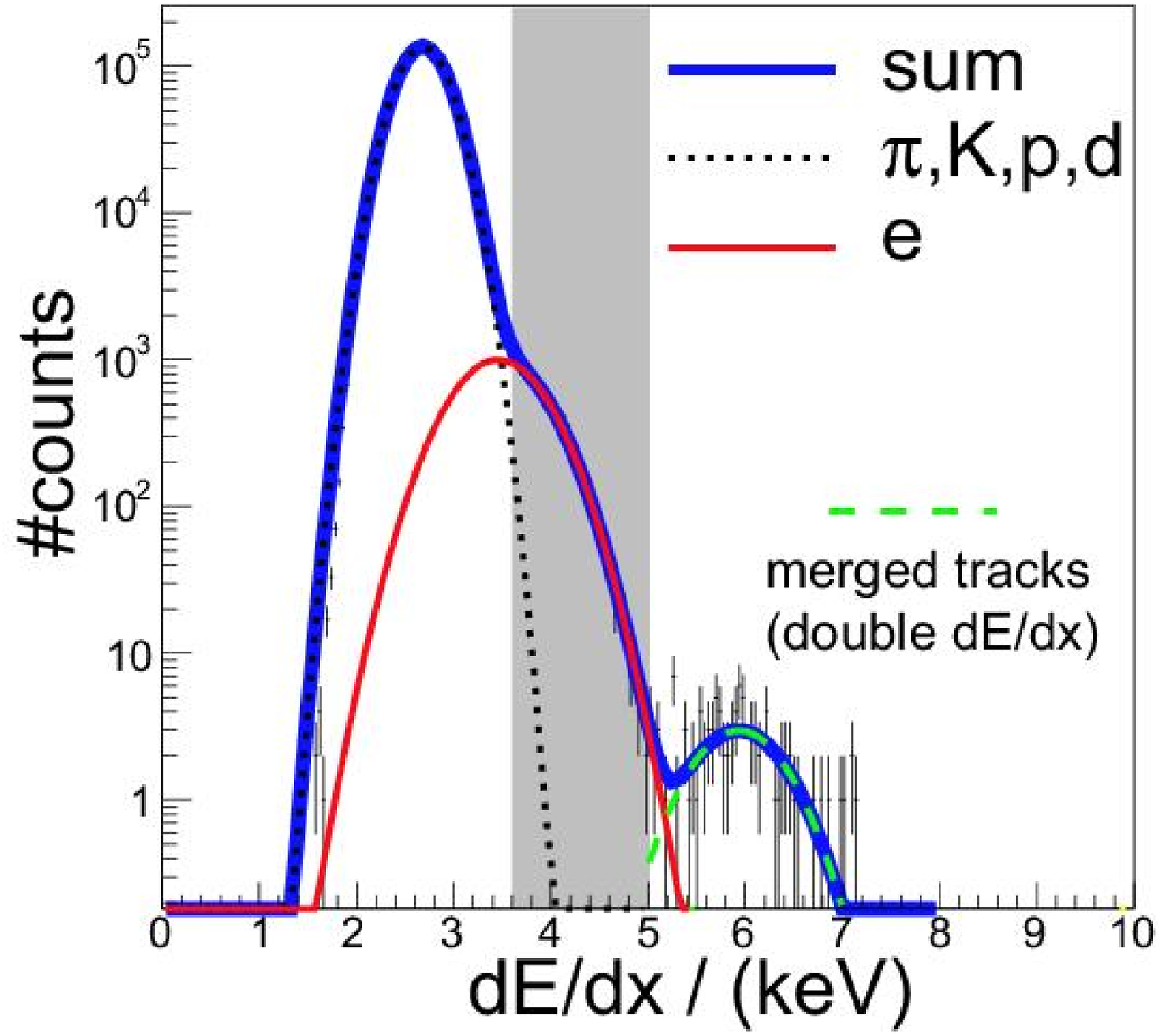}
\end{center}
\caption{(Color online) Left: The specific ionisation of the TPC gas as a function of $log_{10}(p/(GeV/c))$. Right: Distribution of the specific ionisation of the TPC gas in the momentum region $2<p/(GeV/c)<3$.  The three gaussian curves show the contribution of $\pi$, K, p and d centered at about $2.6$~keV/cm, the contribution of electrons at about $3.5$~keV/cm, and background from merged tracks at about $6$~keV/cm.}
\label{fig:dEdx}
\end{figure}

Similar as described in \cite{StarPhysRevLett86_402}, the event-by-event reaction-plane was determined by calculating the Q-vector from charged particles within the phase space region $0.1< p/(\mbox{GeV/c}) <2$  and $|\eta|<0.8$.
% and fulfilling several track quality criteria 
%In additon, the particles' tracks are required to fulfill the following quality criteria:
%
%\begin{itemize}
%\item reconstructed out of at least 15 TPC points (45 being the maximum),
%\item the ratio of TPC points to the maximal possible number of points for a track in a given phase-space region is $> 50\%$,
%\item the helix fit  used for track extrapolation and momentum determination converges with  $0.5<\chi^2/\mbox{ndf}<1.3$,
%\item extrapolates back to the primary vertex within $\pm 2$~cm. 
%\end{itemize}
%
For the electron candidate tracks to be correlated with the reaction-plane 
%were required to fulfill the same quality criteria, however, their 
the momentum range was restricted to $2 < p/(\mbox{GeV/c})<3$ and a cut on the track's specific ionization of the TPC gas was placed at $3.6<dE/dx/(\mbox{keV/cm}) < 5$.  The choice of these  selection criteria is due to the $\pi$, K, p, and deuteron $dE/dx$ bands crossing over the electron band at lower momenta and the limited  separation between the bands above $p=2$~GeV/c as seen in the left panel of Figure \ref{fig:dEdx}. 

The right panel shows the $dE/dx$ distribution in the momentum range $2 < p/(\mbox{GeV/c})<3$. The three gaussian curves represent: a) the combined contributions from $\pi$, K, p, and deuteron tracks centered at about $2.8$~keV/cm, b) the contribution from electrons centered at $\approx 3.5$~keV/cm, and c) background believed from merged tracks\footnote{two tracks close together in the TPC which are falsely identified as one single track} at  $\approx6$~keV/cm.
The shaded areas indicate the $dE/dx$ region accepted, leading to a non-electron contamination of about 10\%, while rejecting about 80\% of the total electron yield. 

\begin{wrapfigure}{r}{9cm}
\begin{center}
\includegraphics[width=9cm]{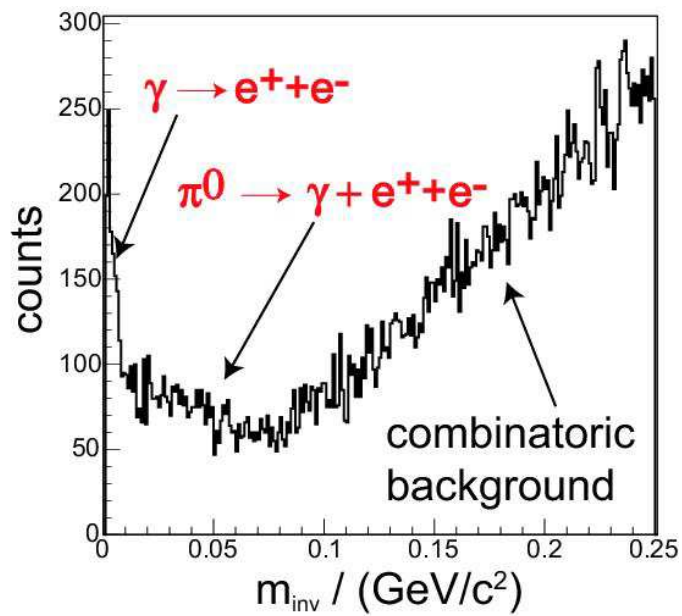}
\end{center}
\caption{Electron removal via $e^+e^-$ invariant mass  method.}
\label{fig:gammaMass}
\end{wrapfigure}

Before these electron candidates are correlated with the reaction-plane, we perform an invariant mass calculation for each candidate with all other tracks within the same event. The so obtained invariant mass distribution (see Figure \ref{fig:gammaMass}) features three distinct regions: a) a sharp peak from $\gamma \rightarrow e^+ + e^-$ conversion in the detector material, b) a smeared out peak from the $\pi^0 \rightarrow e^+ + e^- + \gamma$ Dalitz decay where the $\gamma$ is not reconstructed, and c)  the combinatoric background of random $e^+,e^-$ pairs. All electrons which form a pair with an invariant mass of less that $100$~MeV/c$^2$ are removed from the candidate sample. 

With detailed simulations of the STAR detector and using the $\pi^0$ distribution from \cite{PhenixPhysRevLett91_072301}, we find that this method eliminates about 50\% of all electrons originating from $\pi^0$ decays and $\gamma$-conversions.  In the momentum interval  $2< p/(\mbox{GeV/c})<3$, we estimate our electron sample to constitute to 63\% from D-meson decays and to 37\% from electrons from the remaining photonic sources ($\gamma$ conversion and $\pi^0$ Dalitz decays).  Electron backgrounds originating from other sources, e.g. the $\eta$ Dalitz decay, are expected to be relatively small \cite{StarOpenCharmPaper} and are neglected for this analysis. 

\begin{figure}[htp]
\begin{center}
\includegraphics[width=10cm]{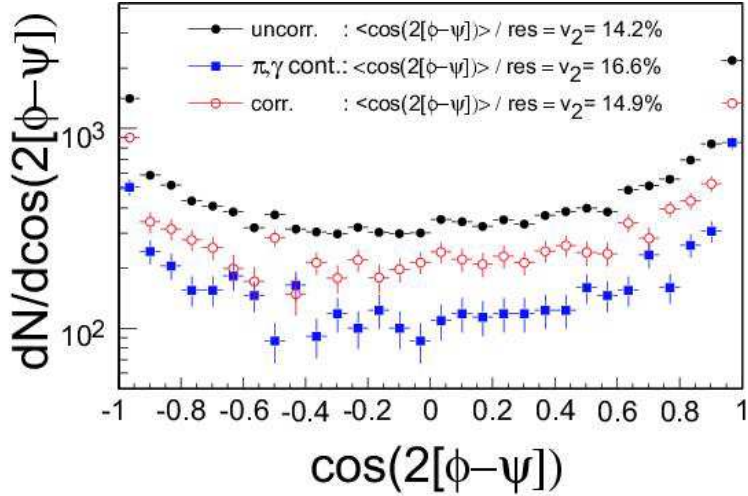}
\end{center}
\caption{(Color online) $\cosTwoPhi$ distribution for uncorrected (solid bullets), corrected (open bullets) and background (squares) $e^\pm$ }
\label{fig:flowCos2Phi}
\end{figure}

The azimuth angles $\phi$ from the remaining $\e^\pm$ are now correlated with the reaction-pane angles $\psi_r  = atan2 \left( Q_y,Q_x \right) $ and the  $\cosTwoPhi$ distribution (see Figure \ref{fig:flowCos2Phi}) recorded. The $v_2$ value can then be calculated $v_2 = \mean{\cosTwoPhi} / \psi^{res}_{RP}$  . Here, $\psi^{res}_{RP} \approx 0.7$  is the reaction-plane resolution which is determined via the sub-event method as described in \cite{StarPhysRevC66_034904}.
\section{Results}
In order to correct for the 37\% $e^\pm$ background from photonic sources,  we again use the $\pi^0$ spectrum from \cite{PhenixPhysRevLett91_072301} onto which we impose elliptic flow according to Figure \ref{fig:scaledV2} ($v_2^{max} = 17\%$).  Passing the resulting azimuthal anisotrop $\pi^0$--distribution through our detector simulation and analysis code, we obtain a $\cosTwoPhi$ distribution as shown by the squares in Figure \ref{fig:flowCos2Phi}. Subtracting this distribution from the  uncorrected $\cosTwoPhi$ distribution (solid bullets),  we obtain the $\cosTwoPhi$ distribution for $e^\pm$ from D-meson decays (open bullets). Again, the elliptic flow amplitude is calculated as  $v_2 = \mean{\cosTwoPhi} / \psi^{res}_{RP}$  and evaluates to $v_{2\e^\pm} \approx 0.15\pm0.02(stat)$  in the momentum interval    $2 < p/(\mbox{GeV/c})<3$.  Our preliminary estimate of systematic uncertainties originates from uncertainties in the $v_{2\pi^0}$ and the exact amount of photonic $\e^\pm$ background and adds up to  $\approx 25\%$ (relative). Other sources of systematic uncertainties (e.g. a slight systematic bias against central events when removing $\gamma$--conversion electrons) are still under investigation.

\begin{figure}[htp]
\begin{center}
\includegraphics[width=10cm]{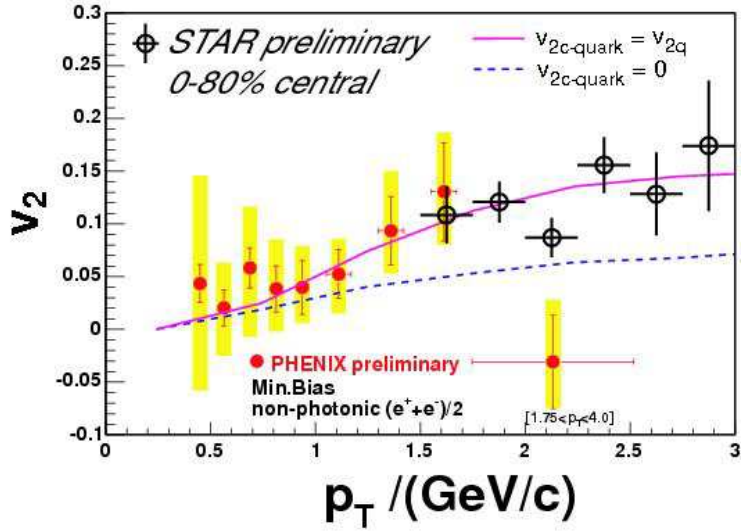}
\end{center}
\caption{(Color online) $v_2$ from non-photonic $\e\pm$ vs transverse momentum in Au+Au collisions at $\sqrt{N_{NN}}=200$~GeV. The open bullets represent preliminary data from this analysis, the full bullets represent results from \cite{KanetaNuclEx0404014}. The solid(dashed) line show the expectations from a quark coalescence model with thermalized and flowing(non-interaction and  azimuthal isotrop) $c$ quarks.  
}
\label{fig:results}
\end{figure}

Figure \ref{fig:results} shows our preliminary single electron elliptic flow results as a function of transverse momentum (open bullets). Again, the vertical bars represent only the statistical uncertainties and our preliminary systematic uncertainty estimate is about $25\%$. The figure also features data from a similar analysis presented by the Phenix experiment \cite{KanetaNuclEx0404014} (full bullets). Neglecting the difference in the centrality of the event samples used (Phenix: min. bias, this analysis 0-80\% central),  our data agree well with the Phenix results and extend them smoothly to higher momenta.

The data points seem to favor the prediction from the quark coalescence model \cite{GrecoNuclTh0405040, GrecoNuclTh0312100} under the assumptions of a partonic stage with thermalized and strongly flowing $c$ quarks (solid line). However, with the present understanding of systematic uncertainties, the non-flow assumption can not be ruled out completely.  

\ack
We thank the RHIC Operations Group and RCF at BNL, and the
NERSC Center at LBNL for their support. This work was supported
in part by the HENP Divisions of the Office of Science of the U.S.
DOE; the U.S. NSF; the BMBF of Germany; IN2P3, RA, RPL, and
EMN of France; EPSRC of the United Kingdom; FAPESP of Brazil;
the Russian Ministry of Science and Technology; the Ministry of
Education and the NNSFC of China; Grant Agency of the Czech Republic,
FOM and UU of the Netherlands, DAE, DST, and CSIR of the Government
of India; Swiss NSF; and the Polish State Committee for Scientific 
Research.

\end{document}